\documentclass[twocolumn,showpacs,pra]{revtex4}
\begin{document}

\title{Violation of Bell inequalities through the coincidence-time loophole}
\author{Peter Morgan}
\email{peter.w.morgan@yale.edu}
\affiliation{Physics Department, Yale University, New Haven, CT 06520, USA.}
\homepage{http://pantheon.yale.edu/~PWM22}

\begin{abstract}
The coincidence-time loophole was identified by Larsson \& Gill (\textit{Europhys. Lett.} \textbf{67}, 707 (2004));
a concrete model that exploits this loophole has recently been described by De Raedt \textit{et al.}
(\textit{Found. Phys.}, to appear).
It is emphasized here that De Raedt \textit{et al.}'s model is experimentally testable.
De Raedt \textit{et al.}'s model also introduces contextuality in a novel and classically more natural way than
the use of contextual particle properties, by introducing a probabilistic model of a limited set of degrees of
freedom of the measurement apparatus, so that it can also be seen as a random field model.
Even though De Raedt \textit{et al.}'s model may well contradict detailed Physics, it nonetheless provides
a way to simulate the logical operation of elements of a quantum computer, and may provide a way forward for more
detailed random field models.
\end{abstract}


\newcommand\iint{{\int\hspace{-0.6em}\int}}

\maketitle

\setlength{\baselineskip}{1.3\baselineskip}

De Raedt \textit{et al.}\cite{RRMKM1,RRMKM2,ZRM} (hereinafter RRMKM) construct a computer model that violates
Bell-type inequalities, which can be used to simulate elements of a quantum computer at
an event by event level.
Although the RRMKM model can be understood as a computing simulation, \emph{not} as a Physics model,
it is a local model that can be said to exploit the ``coincidence-time'' loophole\cite{SL1}, which was 
identified by Larsson and Gill as ``significantly more damaging than the well-studied detection problem''\cite{LG}.
The RRMKM model is more concrete and less general than the models discussed in \cite{LG}.
Such models would be of little interest to most Physicists were it not for the fact that the RRMKM model,
if it is considered as a Physics model, is experimentally testable, and is a prototype for more detailed
random field models.

The coincidence of events is part of the conventional definition of 2-particle states in quantum mechanics:
if we observe two events at time-like separation, they may or may not be caused by the same particle; if we observe
two events at space-like separation, they cannot be caused by a single particle, there must be a 2-particle state (or
a 3-particle state, ...).
For an archetypal experiment that measures a 2-particle state, we may turn to Weihs \textit{et al.}'s
measurement of a violation of Bell inequalities\cite{Weihs}.
In this experiment, two computers recorded the times at which events occurred at each of the two ends of the
experiment, then the two datasets were compared (on a third computer, although this is logically inessential) to
determine when there were approximately matched events, ``Coincidences were identified by calculating time
differences between Alice's and Bob's time tags and comparing these with a time window (typically a
few ns)''\cite[p 5041]{Weihs}.
The storage of two separate datasets followed by subsequent analysis is logically equivalent to a hardware
coincidence circuit, but very usefully allows the retrospective analysis of the coincidences we would have
observed if we had used different hardware coincidence circuits.

The RRMKM model can be understood on two levels: as a computer simulation of individual events; and as a
probabilistic model that captures the properties of the simulation.
The empirical adequacy of an event by event simulation model is established by comparison of statistics of the
computer generated events with statistics of experiments, no reference to a probabilistic model is necessary, so
a simulation approach to Physics is not necessarily parasitic on a probabilistic approach, but the presentation
of computer simulations is not yet, to this author, transparent as Physics.
Here, only the probabilistic approach will be discussed, because it is more appropriate as a conventional
Physics model.

In brief, the RRMKM model understood in a probabilistic way depends on a probability density of the time
at which a single event is observed, $p(t|\mathbf{a}.\mathbf{S})$, having a non-trivial dependence on the
polarization of a light source\cite[\S 6]{ZRM}.
In quantum theory, a rotationally invariant 2-photon quantum state (which we will say --- unrealistically, but for
the sake of simplicity --- has already been determined not to require helical polarization in its description) is 
a mixture of a pure state,
\begin{equation}\label{PureState}
  \rho_p=\psi_p\psi_p^\dagger,\ \psi_p=\frac{1}{\sqrt{2}}\left(\left|H_\mathbf{S}\right>_1\left|V_\mathbf{S}\right>_2-
                              \left|V_\mathbf{S}\right>_1\left|H_\mathbf{S}\right>_2\right),
\end{equation}
which is invariant under rotation of the polarization vector $\mathbf{S}$, and a rotationally invariant mixed state,
\begin{equation}\label{MixedState}
  \rho_m=\frac{1}{2\pi}\int d\mathbf{S} \left|H_\mathbf{S}\right>_1\left|V_\mathbf{S}\right>_2
     \left<V_\mathbf{S}\right|_2\left<H_\mathbf{S}\right|_1.
\end{equation}
Characterization of an optical source requires us to determine a range of such mixtures that model the source to
a chosen empirical accuracy.
A dependence of $p(t|\mathbf{a}.\mathbf{S})$ on the polarization of the light source, if observed, reduces our
ability to limit the range of such states that are empirically adequate.
One conclusion of this letter is therefore that the description of any experiment that depends on coincidences
of events for different polarizations should include a characterization of the dependence of detector delay
on different polarizations, because future experimentalists will have to reproduce that dependence to obtain
the same results.

The RRMKM model understood in a probabilistic way works by constructing a familiar separable hidden variable
model in terms of polarization vectors $\mathbf{S}_1$ and $\mathbf{S}_2$,
\begin{widetext}
\begin{equation}
  p(x_1,x_2,t_1,t_2|\mathbf{a}_1,\mathbf{a}_2)=\frac{1}{4\pi^2}\iint d\mathbf{S}_1d\mathbf{S}_2
            p(x_1|\mathbf{a}_1.\mathbf{S}_1)p(t_1|\mathbf{a}_1.\mathbf{S}_1)
            p(x_2|\mathbf{a}_2.\mathbf{S}_2)p(t_2|\mathbf{a}_2.\mathbf{S}_2)p(\mathbf{S}_1,\mathbf{S}_2),
\end{equation}
where $x_1,x_2\in\{-1,+1\}$, depending on which detector triggers behind a polarizing filter aligned at angles
$\mathbf{a}_1,\mathbf{a}_2$, respectively, at the two ends of the experiment.
To model coincidence as it is described in the Weihs \textit{et al.} experiment, we suppose that $|t_2-t_1|$ must
be less than a length of time $W$, and integrate over all time, to obtain
\begin{equation}
  p(x_1,x_2|\mathbf{a}_1,\mathbf{a}_2)=
      \frac{\displaystyle \iint d\mathbf{S}_1d\mathbf{S}_2
            p(x_1|\mathbf{a}_1.\mathbf{S}_1)p(x_2|\mathbf{a}_2.\mathbf{S}_2)
            w(\mathbf{a}_1.\mathbf{S}_1,\mathbf{a}_2.\mathbf{S}_2,W)p(\mathbf{S}_1,\mathbf{S}_2)}
           {\displaystyle \iint d\mathbf{S}_1d\mathbf{S}_2
                   w(\mathbf{a}_1.\mathbf{S}_1,\mathbf{a}_2.\mathbf{S}_2,W)p(\mathbf{S}_1,\mathbf{S}_2)},
\end{equation}
where the weight function $w(\mathbf{a}_1.\mathbf{S}_1,\mathbf{a}_2.\mathbf{S}_2,W)$ is
\begin{eqnarray}
  w(\mathbf{a}_1.\mathbf{S}_1,\mathbf{a}_2.\mathbf{S}_2,W)&=&
      \iint dt_1dt_2 p(t_1|\mathbf{a}_1.\mathbf{S}_1)p(t_2|\mathbf{a}_2.\mathbf{S}_2)\Theta(W-|t_2-t_1|)\label{Wexact}\\
     &=& 2W\int dt p(t|\mathbf{a}_1.\mathbf{S}_1)p(t|\mathbf{a}_2.\mathbf{S}_2) +O(W^2).\label{Wapprox}
\end{eqnarray}
\end{widetext}
Eq. (\ref{Wexact}) is an integral on a line of width $W\sqrt{2}$, centered on $t_1=t_2$, leading to Eq. (\ref{Wapprox})
when $W$ is small.
With an appropriate choice of $p(t|\mathbf{a}.\mathbf{S})$, $p(x_1,x_2|\mathbf{a}_1,\mathbf{a}_2)$ is not separable
and may violate Bell inequalities\cite[\S 6]{ZRM}, so a local model can be constructed that reproduces the
logical operation of a quantum computer (the logical operation of a quantum computer being independent of the
detailed Physics, the usefulness of this approach will to some extent survive if further experiment invalidates
them as Physics models).
In particular, \cite[\S 6]{ZRM} uses a pseudo-random model for the polarizer that reproduces Malus law, for which
\begin{equation}\label{pxMalus}
  p(x|\mathbf{a}.\mathbf{S})=\frac{1-x}{2}+x(\mathbf{a}.\mathbf{S})^2=\frac{1}{2}(1+x\cos{2\zeta}),
\end{equation}
where $\cos{\zeta}=\mathbf{a}.\mathbf{S}$, introduces a uniform distribution ansatz,
\begin{eqnarray}
  p(t|\mathbf{a}.\mathbf{S})&=&\frac{\Theta(t)\Theta(T(\mathbf{a}.\mathbf{S})-t)}{T(\mathbf{a}.\mathbf{S})},\cr
 \mathrm{where}\ T(\mathbf{a}.\mathbf{S})&=&
     T_0\left[4(\mathbf{a}.\mathbf{S})^2(1-(\mathbf{a}.\mathbf{S})^2)\right]^{d/2}\cr
                         &=&T_0\left|\sin{2\zeta}\right|^d,
\end{eqnarray}
and chooses $p(\mathbf{S}_1,\mathbf{S}_2)$ so that $\mathbf{S}_1$ and $\mathbf{S}_2$ are orthogonal, to construct
a model that matches the predictions of quantum theory for the pure rotationally invariant state of
Eq. (\ref{PureState}) when $d=4$ and $W\ll T_0$ is small.
When $d=0$ or $W>T_0$ is large, the model satisfies Bell inequalities.
The uniform distribution ansatz of this model is inessential to the violation of Bell inequalities, and
the weight function $w(\mathbf{a}_1.\mathbf{S}_1,\mathbf{a}_2.\mathbf{S}_2,W)$ does not determine
$p(t|\mathbf{a}.\mathbf{S})$, so there is an infinite class of functions that violate Bell inequalities.

The RRMKM model is experimentally testable, provided we assume that after passing through a polarization filter
the unobservable polarization vector $\mathbf{S}$ is aligned with the polarization filter (this assumption is
required in conjunction with Eq.(\ref{pxMalus}) to reproduce Malus law; the simplest quantum mechanical modeling of
polarization filters is, comparably, as a projection of the state to the same alignment as the polarization filter).
We can measure the delay dependency function $p(t|\mathbf{a}.\mathbf{S})$ for a given detector by
directing a source of known polarization through a briefly open gate and observing whether there are
timing differences for different relative orientations.
Then we can compute the resulting weight function,
$w(\mathbf{a}_1.\mathbf{S}_1,\mathbf{a}_2.\mathbf{S}_2,W)$,
which establishes how much violation of the Bell inequality can be accounted for by detection delay.
It is plausible from the classical optics of crystals that there will be such timing
dependencies\cite[\S 14.3.2,\S 14.4.1]{BW}, but the extent of the timing dependencies required is considerable,
since the violation of the Bell inequalities in Weihs \textit{et al.}'s experiment only diminishes to zero as
$W$ is increased beyond 300ns\cite[Fig. 3]{ZRM}, corresponding to an effective path length difference of 100m.
If the whole violation of the Bell inequalities by an experiment can be accounted for by the delay dependency of
the detectors used, then there is a sense in which we have so far failed to characterize the state of the light
source as definitely $\rho_p$, definitely $\rho_m$, or as one of the continuum of intermediate mixtures.
Provided we consistently use detectors that have the same delay dependency and we use the same procedure to
determine event coincidences, however, we can continue to use $\rho_p$ to describe a light source (supposing that
the measurement results --- on the (false) assumption that we are using detectors for which
$p(t|\mathbf{a}.\mathbf{S})$ is independent of $\mathbf{a}.\mathbf{S}$ --- support $\rho_p$ as a model), but
if we use different detectors or different event coincidence criteria we must reassess the empirical effectiveness
of $\rho_p$.
If $\rho_m$ successfully models an experiment when detectors are modelled accurately, this of course does not mean
that an experiment \emph{is} classical.

If detectors generally prove to have nontrivial dependencies of $p(t|\mathbf{a}.\mathbf{S})$ on
$\mathbf{a}.\mathbf{S}$, we can retain a relatively straightforward conceptual position by insisting that
an \emph{ideal} quantum detector has no such dependency.
This is a reasonable position to adopt even if all detectors have nontrivial dependencies on
$\mathbf{a}.\mathbf{S}$, provided it can be proven that there is no limit to the reduction of such dependencies.
Dependence of $p(t|\mathbf{a}.\mathbf{S})$ on $\mathbf{a}.\mathbf{S}$ in both the detectors is
an \emph{engineered} nonlocal correlation between the measurement apparatuses, because the detectors have the
same internal structure.
Together with the nonlocally defined determination of coincidence of events at space-like separation, this
nonlocal correlation is enough to introduce a nondynamical nonlocality into this classical model.
Seen in this way, the construction of the experimental apparatus is an example of what is pejoratively termed
``conspiracy''.

Modeling the dependence of $p(t|\mathbf{a}.\mathbf{S})$ on $\mathbf{a}.\mathbf{S}$ effectively
introduces a small, critical part of the experimental apparatus into the experimental model, so we should also
understand the model to be ``contextual''.
The contextuality of the RRMKM model accommodates the Fine and Accardi discussion of Bell
inequalities\cite{Accardi,Fine,deMuynck,Laudan,Streater,HessPhilipp}, without, however, introducing contextual
particle properties, which are rightly anathema to classical particle physics.
Noting that the discussion of classical models for quantum mechanical systems has always stalled on whether the
necessary modifications of classical physics are natural rather than whether they are possible, the RRMKM model
moves classical models one more step towards naturalness.
Details of the experimental apparatus have a subtle impact on how we model the experiment, making it increasingly
difficult to understand the experiment as a ``measurement'' of ``the system we are measuring''.
As we consider experiments in progressively more detail, we are forced to introduce more details of the
experimental apparatus, so we cannot confine our quantum models to small numbers of electrons and photons,
with the experimental apparatus represented only by a measurement operator.
In effect, as we include more of the experimental apparatus, and increasingly finer details, we move the Heisenberg
cut outwards into the world, making our experiments more and more capable of being modeled by classical random
field methods\cite{Morgan1,Morgan2}.
The significance of the RRMKM model is that it moves a \emph{small} characteristic property of the measurement apparatus
into our detailed discussion of the measurement, without requiring a perfect description of the whole measurement device.
Thanks to the arguments in \cite{Morgan1}, we know that if we move \emph{enough} of the measurement apparatus into
a model for an experiment, a classical model can violate Bell inequalities.
Even if the RRMKM model is ruled out by measurement of $p(t|\mathbf{a}.\mathbf{S})$, nonetheless it gives a novel
way to introduce contextuality into classical models, in a classically acceptable way, without introducing
contextual particle properties.

Note that the above discussion is not affected by the critique of Hess and Philipp's discussion\cite{HP1,HP2} by
Gill \textit{et al.}\cite{Gill1,Gill2}.
Here, timings and coincidences of events are explicitly at issue in a different way than highlighted in those papers,
as noted by Larsson and Gill\cite{LG}.

Quantum mechanics is not under threat as an engineering discipline --- it is much too useful to separate the
world into ``the measurement apparatus'' and ``the system that is measured''; into ``the measurement apparatus'' and
``the preparation apparatus''; or even into ``the Universe'' and ``ideal quantum measurement''; all of which allow
the mathematical tools of quantum theoretical measurement operators and Hilbert spaces of states to be used.
This splitting of the world into two parts seems to be always possible For All Practical Purposes, \emph{but} the choice
is always pragmatic.
This is of course the arbitrariness of the Heisenberg cut: the way in which we split the world into two parts is not
a perfect truth about the world.
This is a \emph{fundamental} limitation of the mathematical tools of quantum theory.
The immediate alternative, however, a classical holistic model that explicitly models quantum fluctuations\cite{Morgan1,Morgan2},
is \emph{no} better, since there cannot be a model of the whole universe: there is always a separation of the world
into what is in the model and what is not in the model, so classical modeling is equally pragmatic.
I therefore make no claims that we should construct classical models, but, pragmatically, it might sometimes be useful
to do so, and we can better understand quantum (field) theory by better understanding the relationship between
quantum and classical models.
In particular, as quantum theoretical models increasingly introduce more details of the measurement apparatus, in a
constant pursuit of more accuracy, contextual classical models will increasingly become alternatives of comparable
complexity.

I am grateful to Hans De Raedt for correspondence and for clarifications of details of the RRMKM model.

\end{document}